\documentclass{ar-1col}
\usepackage[numbers,sort&compress]{natbib}

\usepackage{url}
\usepackage{natbib}
\usepackage{textcomp}
\usepackage{color,ulem}

\usepackage{textcomp}   
\usepackage{mhchem}
\usepackage{bm}

\jname{Xxxx. Xxx. Xxx. Xxx.}
\jvol{AA}
\jyear{YYYY}
\doi{10.1146/((please add article doi))}

\begin{document}

\markboth{Mozur et al.}{Dynamics in Hybrid Perovskites

}

\title{Cation Dynamics in Hybrid Halide Perovskites}

\author{Eve M. Mozur,$^1$ James R. Neilson,$^1$
\affil{$^1$Chemistry Department, Colorado State University, Fort Collins CO, USA, 80523; email: jrn@colostate.edu}}

\begin{abstract}
Hybrid halide perovskite semiconductors exhibit complex, dynamical disorder while also harboring  properties ideal for optoelectronic applications that include photovoltaics.  However, these materials are structurally and compositionally distinct from ``traditional'' compound semiconductors composed of tetrahedrally-coordinated elements with an average valence electron count of silicon.  As discussed here, the additional dynamic degrees of freedom of hybrid halide perovskites underlie many of their potentially transformative physical properties.  Neutron scattering and spectroscopy studies of the atomic dynamics of these materials have yielded significant insights to the functional properties.  Specifically, inelastic neutron scattering has been used to elucidate the phonon band structure, and quasi-elastic neutron scattering (QENS) has revealed the nature of the uncorrelated dynamics pertaining to molecular reorientations.   Understanding the dynamics of these complex semiconductors has elucidated the temperature-dependent phase stability and origins of the defect-tolerant electronic transport from the highly polarizable dielectric response.  Furthermore, the dynamic degrees of freedom of the hybrid perovskites provides additional opportunities for application engineering and innovation.  
\end{abstract}

\maketitle

\tableofcontents

\section{INTRODUCTION: HYBRID HALIDE PEROVSKITE SEMICONDUCTORS}

Hybrid halide perovskites and their structural derivatives have advantageous optoelectronic properties for a relatively wide range of applications. The tunable band gap, defect tolerance and long electronic excited state lifetimes of these perovskites make them ideal semiconductors for photovoltaics \cite{kim_high-efficiency_2020,stranks_metal-halide_2015}. The heavy elements in these materials also allows their use in X-ray and $\gamma$-ray  detection \cite{xu_detection_2017, nazarenko_single_2017}.  Changing the composition or reducing the effective dimensionality changes the electronic structure and  exciton binding energy to increase the efficiency of light emission;   several materials even reemit white light intrinsically \cite{smith_diversity_2018}. This family of materials provides new opportunities for building the structure-property relationships of next-generation semiconductors. 

\begin{figure}[b]
    \centering
    \includegraphics[width=6.3in]{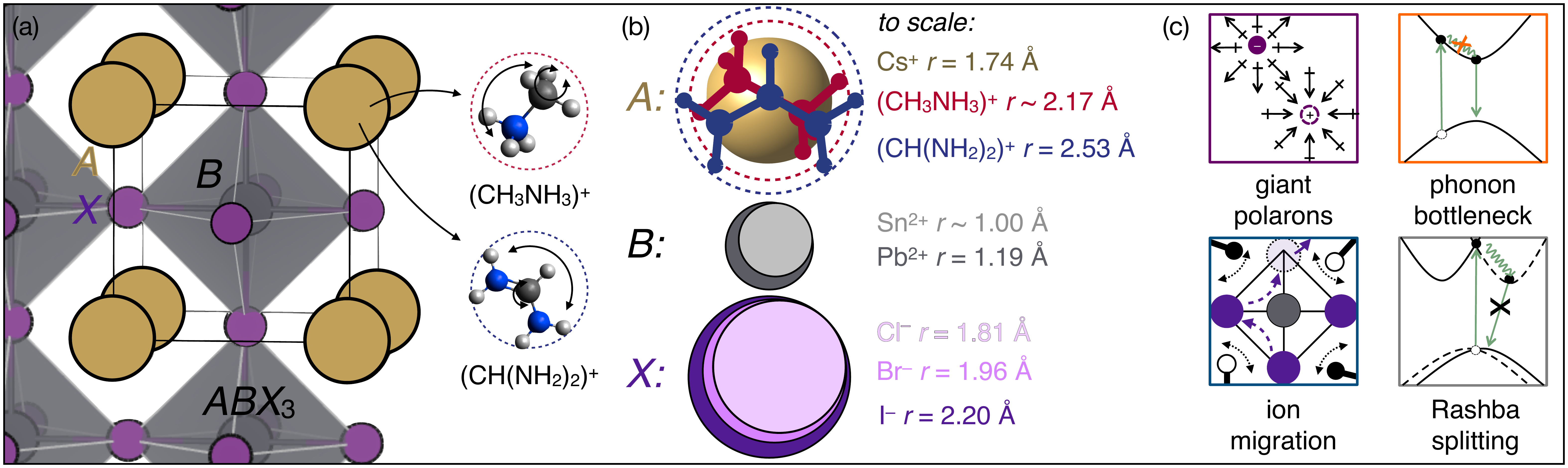}
    \caption{(a) Schematic illustration of the plastic crystalline nature of hybrid perovskites of the formula, $ABX_3$, as well as the principal rotations of methylammonium, \ce{CH3NH_3^+}, and formamidinium, \ce{CH(NH2)_2^+}.  
    (b) Cartoon illustrating the scale of the various ions found in hybrid halide perovskites based upon their Shannon-Prewitt radii \cite{shannon_revised_1976} or estimated molecular size \cite{Kieslich2015}.  
    (c) Commonly proposed electron-phonon interactions in hybrid halide perovskites involving organic cation dynamics, as described in the text.
        }
    \label{fig:structure}
\end{figure}

Hybrid  halide perovskites exhibit additional dynamic degrees of freedom when compared to traditional compound semiconductors (e.g., CdTe, GaAs, TlBr). Materials such as \ce{(CH3NH3)PbI3} and \ce{(CH(NH2)2)PbI3} are plastic crystals, as schematically depicted in \textbf{ Figure~\ref{fig:structure}(a)}: the organic cations exhibit translational symmetry with respect to their center of mass but without  orientational  order\cite{staveley_phase_1962}. The molecular `sphere' approximated by rotating the molecules can be used to compute a revised Goldschmidt tolerance factor based on the ionic radii of the constituent ions (\textbf{Figure~\ref{fig:structure}(b)}) \cite{Kieslich2015}, as often used to predict the formation of a crystal structure analogous to \ce{CaTiO3}, the mineral  Perovskite \cite{helen_d_megaw_crystal_1973}.  Similar to oxide perovskites,  the halide perovskites typically undergo phase transitions on cooling, as characterized by a reduction in symmetry from cubic to tetragonal to orthorhombic related to cooperative tilting of the octahedral framework \cite{mitzi_templating_2001,Woodward1997}. However, the rich and subtle details of these transitions and the resulting properties of each phase cannot be fully understood without considering the plastic crystalline nature of these materials \cite{ghosh_polarons_2020}.

Various advantageous electronic properties in hybrid perovskites have been attributed to the dynamic, plastic crystalline behavior.  While long electronic excited state lifetimes have been linked to collective dynamics of the extended octahedral framework (i.e., phonons) \cite{even_analysis_2014,neukirch_polaron_2016, herz_charge-carrier_2017, freundlich_carrier_2016, wright_electronphonon_2016}, such long-lived excited states are also linked to  organic cation reorientational dynamics  by way of giant polaron \cite{miyata_lead_2017}, phonon bottleneck \cite{yang_observation_2015}, or Rashba splitting \cite{kim_switchable_2014} based mechanisms  (\textbf{Figure~\ref{fig:structure}(c)}; described in more detail in Section~\ref{sec:dipoles}). Furthermore, the presence of an organic cation has been associated with the ``self healing'' of photoinduced defects \cite{ceratti_self-healing_2018} as well as in aiding in photoinduced ionic conductivity \cite{eames_ionic_2015,kim_large_2018,senocrate_nature_2017}. The ionic mobility of hybrid halide perovskites is often linked to decomposition \cite{wang_review_2019,zhou_cation_2019,deretzis_stability_2018,nagabhushana_direct_2016} and is hypothesized to relate to stationary atomic fluctuations \cite{lai_intrinsic_2018},  akin to the ``paddlewheel'' mechanism in materials studied for alkali ion conduction \cite{zhang_coupled_2019,sun_rotational_2019}.
Similar  dynamics are thought to influence relaxation pathways in layered perovskite derivatives of interest as phosphors \cite{smith_structural_2017, gong_electron-phonon_2018}; the microscopic details remain an active area of research.  
Regardless of the proposed mechanism, the impressive ``defect tolerant'' properties of hybrid halide perovskites advance high-efficiency photovoltaics \cite{steirer_defect_2016,brandt_identifying_2015,kim_high-efficiency_2020}. This review focuses on how neutron scattering, particularly quasi-elastic neutron scattering, has addressed questions related to these properties by characterizing  the dynamic nature of organic cations in hybrid perovskite semiconductors.

\begin{marginnote}
 \entry{Phonon Bottleneck }{A bottleneck of heat-transporting phonons prevents electronic relaxation, such as the relaxation of ``hot'' carriers to the band edges \cite{potz_hot-phonon_1987}.}
 \entry{Rashba splitting }{Broken inversion symmetry in the presence of strong spin-orbit coupling splits doubly degenerate bands into spin-up vs spin-down bands in momentum space to yield a material with close direct and indirect transitions, enabling strong absorption but long-lived carriers.}
\end{marginnote}

\begin{figure}[t]
\includegraphics[width=6.3in]{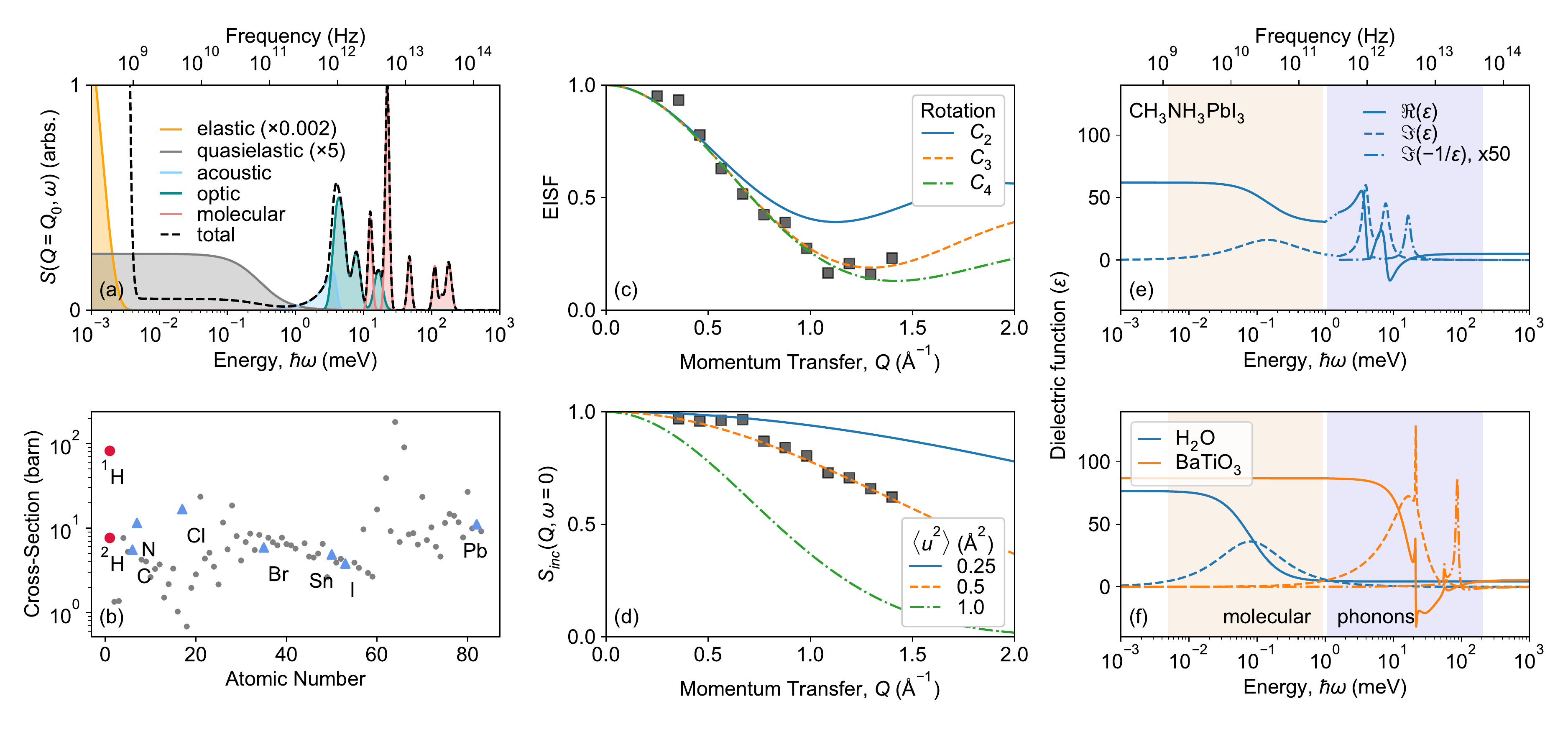}
\caption{(a) Schematic representation of features observable in neutron spectroscopy, including elastic scattering from static atoms, quasi-elastic neutron scattering (QENS) from molecular reorientational dynamics, and inelastic neutron scattering from quantized lattice vibrations (acoustic and optical phonons) and molecular vibrations. (b) Neutron scattering cross section of different natural-abundance averaged elements \cite{sears_neutron_1992}, highlighting the large cross section of \ce{^1H} relative to the other elements germane to hybrid perovskites.  
(c) The elastic incoherent structure factor (EISF) used to extract details of molecular motion from momentum transfer dependence of the QENS, illustrated by three different jump models corresponding to $C_2$, $C_3$, $C_4$ rotations of the same hypothetical molecule. Grey squares illustrate hypothetical data at discrete values of $Q$ from one sample at fixed temperature.  Different jump models can be challenging to discriminate without a sufficient $Q$-range.
(d) Schematic illustration of the incoherent elastic scattering as a function of momentum transfer, $S_{inc}(Q,\omega=0)$, used to extract the neutron scattering cross-section weighted atomic mean-squared displacement, $\langle u^2 \rangle$ = 0.5 \AA$^2$. Grey squares illustrate hypothetical data at discrete values of $Q$ from one sample at fixed temperature.
(e,f) Dielectric function, $\epsilon$, including the real (solid line), imaginary (dashed), and $\Im(-1/\epsilon)$ (dotted) components of (e) \ce{CH3NH3PbI3} \cite{anusca_dielectric_2017,sendner_optical_2016} at 298 K as compared to (f) molecular, liquid \ce{H2O} at 298~K \cite{ellison_permittivity_2007}  and paraelectric ceramic \ce{BaTiO3} at 1000~K \cite{luspin_soft_1980}. The frequency-dependent dielectric function highlights contributions from either molecules (beige shading) or phonons (lavender shading), as detectable by (a) neutron scattering.  
}
\label{fig:neutron}
\end{figure}

\begin{textbox}[ht]\section{NEUTRON SCATTERING, A BRIEF PRIMER}

Neutron scattering is a powerful tool to investigate the structure and dynamics of materials through analysis of the dynamic structure factor, $S(\bm{Q},\omega)$ (\textbf{Figure~\ref{fig:neutron}(a)}). The large mass of neutrons, compared to photons or electrons, means that neutrons have a relatively large momentum that allows them to couple strongly to atomic and molecular dynamics. The scattering cross section depends on  interactions of the neutron with the nucleii within each atom  and is therefore isotope dependent (\textbf{Figure~\ref{fig:neutron}(b)},\cite{sears_neutron_1992}). The lack of monotonic trends in cross section with atomic number also provides complementary contrast to X-ray scattering techniques, and the large (incoherent) cross section of \ce{^1H} makes neutron spectroscopy ideal for studying hybrid perovskites.  The modulation of a neutron's speed through a moderator once released from either a reactor or spallation source allows for tunability of incident energy. Therefore, low wavelengths (0.5 to 4 \AA) are suited for atomic-resolution diffraction, while their energies ($\mu$eV-eV) are ideal for spectroscopy of atomic and molecular dynamics.  
The neutron scattering techniques most relevant for investigating the structure and dynamics of solid state materials include neutron diffraction, quasi-elastic neutron scattering (QENS), and inelastic neutron scattering (INS), as summarized in \textbf{Figure~\ref{fig:neutron}}. Diffraction experiments are not typically energy resolved; therefore, they  characterize predominantly elastic interactions that inform the average structure (intensity of elastic to inelastic scattering events $\sim$100:1).  INS probes the phonon and molecular vibrational density of states between $\sim$5 meV to 1000 meV. QENS probes dynamics on the $\mu$eV energy scale, encompassing molecular reorientations, translations, and diffusion, as described in the next breakout panel.  These dynamics map onto the dielectric response function illustrated in \textbf{Figure~\ref{fig:neutron}(e,f)}.   

Currently, these experiments can only be conducted at large, shared spallation- or reactor-based neutron sources.   As such, INS is complementary to infrared and Raman spectroscopy; QENS is complementary to other techniques sensitive to motions on $10^{-9}$~s to $10^{-12}$~s time scales, such as nuclear magnetic spectroscopy (also nucleus specific) and optical spectroscopies (e.g., Brillouin light scattering \cite{letoublon_elastic_2016} or ultrafast vibrational spectroscopy\cite{bakulin_real-time_2015}).  However, the large neutron scattering cross section of hydrogen makes QENS ideal for studying the molecular dynamics in hybrid halide perovskites.  
\end{textbox}

\begin{textbox}[ht]\section{QUASI-ELASTIC NEUTRON SCATTERING (QENS), A CLOSER LOOK}

Quasi-elastic neutron scattering (QENS) can be performed on neutron spectrometers with  excellent energy resolution ($\Delta E \sim$ 1 to 3 $\mu$eV).   The quasi-elastic scattering (e.g., energy transfer $\neq 0$ within the instrumental resolution) yields information about slow motions corresponding to relaxation of nearly stationary states, such as diffusion or molecular relaxation via translations or rotations.  These relaxations provide a Doppler broadening of the elastic scattering due to the small energy transfer between neutrons and the  moving nuclei in the sample.  This is often differentiated from inelastic scattering, where the neutrons excite or are excited by discrete motions with finite energies (e.g., internal molecular vibrations).  

To analyze the molecular motions, one can use a ``jump model,'' as molecules are assumed to reside in preferred orientations with frequent (e.g., GHz) jumps between symmetry-related (site and molecular) orientations \cite{bee_quasielastic_1988}.  
The quasi-elastic linewidth relates to the timescale of this residence time between jumps (\textbf{Figure~\ref{fig:neutron}(a)}, Lorentzian half width at half maximum  = $\hbar/\tau$). 
However, motions with a frequency faster than the maximum energy range of the spectrometer (anywhere from $\hbar \omega_{max}$ = $\pm$30 $\mu$eV to $\pm$1000 $\mu$eV, depending on instrument configuration) appear as a background and are not detected.  
Since one typically measures the incoherent structure factor, $S_\text{inc}(Q,\omega)$, the  $Q$-dependence of the elastic scattering relative to the sum of elastic and quasi-elastic scattering (i.e., the elastic incoherent structure factor, EISF~$= I_\text{elastic}/(I_\text{elastic} + I_\text{quasielastic})$ )  can be fit with a jump model to extract the type of motion, including: the local symmetry of motion, distance of motion, and geometry of object moving, as illustrated in \textbf{Figure~\ref{fig:neutron}(c)}. 

With such excellent energy resolution on a neutron spectrometer, one can accurately measure the fraction of elastic scattering.  The incoherent elastic scattering yields information about thermal motion, which follows a $Q$-dependent attenuation depending on the mean squared displacement (MSD) of atomic motion via the the Debye-Waller factor = $ \exp(-q^2 \langle u^2\rangle/3)$, for isotropic, harmonic motion, as shown in \textbf{Figure~\ref{fig:neutron}(d)}.  These experiments can be accomplished quickly on some instruments, thus permitting parametric studies, as to follow the temperature dependence of thermal motion through phase transitions.  

\end{textbox}

\section{BACKGROUND: DYNAMICS AND ELECTRONIC PROPERTIES}

The frontier electronic states of halide perovskites couple primarily to the low energy optical phonons. The influence of lattice dynamics on semiconductors typically revolves around a discussion of the scattering of charge carriers from electron-phonon coupling.  In compound semiconductors with the diamond-type structure such as GaAs,  optical phonons  reside near $\sim$40 meV; thus they are not appreciably populated at room temperature and play a relatively small role in electron-phonon scattering processes \cite{pelant_luminescence_2012}.  In contrast, halide perovskite optical phonons reside from $\sim$5 meV to $\sim$12 meV (\textbf{Figure \ref{fig:neutron}}). Therefore, a complete understanding of the charge transport necessitates  characterization of the electron-phonon coupling. In the following section, we provide a brief overview of the importance of these collective dynamics in the context of light-absorbing hybrid semiconductors for photovoltaics, and then we discuss the presence and role of localized (i.e., aperiodic) dynamics and their proposed roles in influencing the optoelectronic performance of hybrid perovskites and related semiconductors.

\begin{marginnote}
\entry{Exciton}{Used here as the Coulombically-attracted bound state of negative and positive charge carriers in a crystal.}
\entry{Fr\"ohlich interaction}{Coupling of a charge to the presence of the electric field generated from a longitudinal optical phonon \cite{yu_cardona}.}
\entry{Polaron}{The distortion of an ionic lattice in response to extra charges, akin to the Fr\"ohlich mechanism. }
\end{marginnote}

\subsection{Collective, Periodic Dynamics: Phonons}

Optical phonons, primarily derived from the cooperative dynamics within the octahedral framework, drive the phase behavior of halide perovskites.  Akin to the ferroelectric perovskite oxides  with symmetry-lowering phase transitions on cooling from a paraelectric phase \cite{megaw_origin_1952}, the phonons related to these phase transitions contribute a significant dielectric response, as in the case of \ce{BaTiO3} (\textbf{Figure \ref{fig:neutron}(f)}).   Perovskite halides undergo similar symmetry-lowering phase transitions involving changes to the octahedral tilting upon cooling, which relate to soft phonon modes \cite{swainson_soft_2015, beecher_direct_2016, weadock_question_2020}. Significant entropy releases accompany these canonical phase transitions \cite{katan_entropy_2018}, as related to the loss of orientational degrees of freedom of the organic molecule within the changing shape of the nominally cuboctahedral $A$-site void
\cite{onoda-yamamuro_calorimetric_1990,onoda-yamamuro_thermal_1991,onoda-yamamuro_dielectric_1992, onoda-yamamuro_neutron-diffraction_1995}.

\begin{marginnote}
\entry{Coherent scattering}{Yields structural and dynamical structural information. }
\entry{Incoherent scattering}{Yields information independent of positional correlations.}
\end{marginnote}

These same collective dynamics influence charge transport primarily through electron-phonon coupling. The optical phonons generate just over half of the polarizability in the dielectric response of \ce{CH3NH3PbI3}, as portrayed in \textbf{Figure~\ref{fig:neutron}(e)}.  In  halide perovskites, electron-phonon scattering via the Fr\"ohlich interaction limits the mobility of both free carriers and excitons; therefore, the Fr\"ohlich interaction is the most important form of electron-phonon coupling \cite{herz_charge-carrier_2017, freundlich_carrier_2016, neukirch_polaron_2016}.  The magnitude of the Fr\"ohlich interaction in \ce{CH3NH3PbI3} ($\alpha = $ 1.7-2.4  \cite{miyata_lead_2017}) derives predominantly from a $\sim$11 meV longitudinal optic (LO) phonon (see \textbf{Figure~\ref{fig:neutron}(e)})  and compares to that of other halide semiconductors, such as TlBr ($\alpha = $2.05); this contrasts with the smaller coupling constants of diamond-lattice, tetrahedral semiconductors GaAs ($\alpha = $0.068) and CdTe ($\alpha = $0.39) \cite{sendner_optical_2016,wright_electronphonon_2016}.

\subsection{Independent, Aperiodic Dynamics}

\subsubsection{Inorganic dipolar dynamics}

The fluctuations of lone pair electrons modify the structure and properties of hybrid perovskites.  They have been linked to many exciting aspects of hybrid perovskites, including the large ionic dielectric responses,  light holes, and unconventional band gap: all contribute to the favorable electronic transport and ``defect tolerance'' \cite{fabini_underappreciated_2020}. These ``inorganic dipoles'' \cite{remsing_new_2020} resemble the behavior of plastic crystals with permanent, molecular dipoles (e.g., potassium cyanide \cite{loidl_orientational_1989}), but with dynamics on a fast $10^{14}$ Hz time scale rather than the slower $\sim10^{12}$ Hz time scale that is associated with the ionic dynamics portrayed in \textbf{Figure~\ref{fig:neutron}(e,f)}.

\begin{marginnote}
\entry{Lone Pair Electrons}{Ge(II), Sn(II), Pb(II), Sb(III), and Bi(III), have the valence electron configuration, ``$ns^2$,'' that can become stereochemically active to create space-filling lone pair electrons \cite{walsh_chemical_2011}, as employed to describe the nearly tetrahedral molecular geometry of \ce{NH3}.}
\end{marginnote}

All-inorganic perovskites exhibit significant polar dynamics from anharmonic ionic fluctuations, attributed to a ``head-to-head'' motion of cesium ions \cite{yaffe_local_2017}. 
Such polar fluctuations and deformations of the $[BX_6]$ octahedra have been implicated in the formation of polarons, which screen excited state charge carriers that increase the excited state lifetimes \cite{wu_light-induced_2017, park_excited-state_2018}. This explains how all inorganic halide perovskites yield comparable transient electronic responses and device performance to their ``hybrid'' relatives \cite{zhu_organic_2017}.

\subsubsection{Organic cation dynamics} \label{sec:dipoles}

Hybrid halide perovskite are plastic crystals, and the dynamically disordered organic cations impart new degrees of freedom relative to conventional compound semiconductors.  These dynamics play a role in various functional properties, deriving primarily from their polarizability \cite{even_molecular_2016}.  Nearly half of the static dielectric polarizability of \ce{CH3NH3PbI3} derives from the dynamic molecules, akin to that of a polar solvent such as water (\textbf{Figure~\ref{fig:neutron}(e,f)}).  This polarizability has several implications for charge transport, summarized in \textbf{Figure~\ref{fig:structure}(c)} and described below.

Although the exact mechanism behind the long excited state lifetimes for thermalized and hot carriers remains debated, the orientational dynamics of the organic molecules play an important role in many (and not mutually exclusive) hypotheses.  Organic cation dynamics have been implicated in the formation of large polarons that interact with both exciton \cite{thouin_phonon_2019} and free charge carriers \cite{wolf_polaronic_2017,frost_slow_2017}.  The presence of dynamic organic cations accelerates the polaron formation time (0.3 ps in \ce{CH3NH3PbBr3} vs 0.7 ps \ce{CsPbBr3} \cite{miyata_large_2017}), which extends the lifetime of hot carriers \cite{zhu_screening_2016}.  The coupling of inorganic lattice distortions and organic cation reorientations has been shown to screen charge carriers and excitons \cite{even_analysis_2014,neukirch_polaron_2016}.    For example, the lack of interactions between independently reorienting molecules yields a low vibrational group velocity, and thus poor thermal conductivity with a high heat capacity -- an ideal recipe for a ``phonon bottleneck,'' in which energetic carriers cannot thermalize efficiently with the lattice due to poor heat removal, as detected from  pump-probe spectroscopy experiments  \cite{yang_observation_2015,yang_acoustic-optical_2017,monahan_room-temperature_2017} and supported by a polaron model backed by density functional theory calculations \cite{frost_slow_2017}. 

\begin{marginnote}
\entry{Plastic Crystals}{Characterized by a lack of orientational order with a preservation of translational order. Potassium cyanide is a canonical example, where cyanide ions are orientationally disorded, but their center of rotation has a well defined position. The dynamic disorder and the subsequent temperature-dependent orientational ordering transitions of plastic crystals give rise to numerous entropic, dielectric, and (ferro)elastic anomalies \cite{staveley_phase_1962,loidl_orientational_1989,hochli_orientational_1990}. }
\end{marginnote}

Furthermore, the presence of polar fluctuations has the potential to locally and transiently break inversion symmetry of the bulk crystal.  Given the slow orientational time scale of the organic dynamics ($\sim 5-15$ ps, \textbf{Figure~\ref{fig:summary}}) relative to electronic dynamics, it has been proposed that Rashba spin-split valence bands can give rise to a direct band gap-like efficient optical absorption transitions\cite{kim_switchable_2014,marronnier_influence_2019,kepenekian_rashba_2015,mosconi_rashba_2017}; carrier thermalization leads to a shifted momenta, thus resulting in an indirect-like recombination and longer lifetime (\textbf{Figure~\ref{fig:structure}(c)}).  Experiments have concluded that the halide perovskites can support static \cite{wang_spin-optoelectronic_2019,glinka_distinguishing_nodate} and dynamic \cite{wu_indirect_2019,niesner_structural_2018,Etienne2016,motta_revealing_2015} Rashba spin splitting, with recent evidence that the static effect only persists at low temperatures in hybrids, and that both all-inorganic and hybrid materials host a dynamic Rashba effect at room temperature \cite{ryu_static_2020}. The dynamic Rashba effect appears to originate from anharmonic, polar thermal fluctuations of the inorganic octahedra that may arise from lone pair electron activity  and their interactions with organic cation reorientations or cesium motions \cite{ryu_static_2020,yaffe_local_2017,wu_light-induced_2017,park_excited-state_2018}. Therefore, the interplay between organic- and inorganic-centered dynamics has the potential to influence the electronic excited state dynamics.  

The dynamic organic molecules also impact the ionic transport and diffusion.  Hybrid halide perovskites have a significant ionic contribution to the conductivity under various circumstances \cite{eames_ionic_2015,frost_what_2016,senocrate_nature_2017,shikoh_ion_2020}, including under light illumination \cite{kim_large_2018}. 
While ion motion leads to device hysteresis  \cite{eames_ionic_2015}, phase separation \cite{slotcavage_light-induced_2016}, and decomposition \cite{kim_large_2018}, motion of the halogens can lead to the creation of short-lived electronic defects.  Such rapid defect creation can lead to defect densities above the thermodynamic limit with sufficiently short lifetimes as to be benign to carrier scattering \cite{rakita_when_2019},  further supported by the self-elimination of intrinsic defects on cooling \cite{chen_self-elimination_2020}.    Furthermore, the timescale of defect motion is faster in hybrid perovskites, as illustrated by the relatively fast self-healing of \ce{CH3NH3PbBr3} and \ce{CH(NH2)2PbBr3} from laser-induced defects relative to \ce{CsPbBr3} \cite{ceratti_self-healing_2018}.  While this leads to a high tolerance to unintentional defects, it makes controlled doping very challenging.  

Given the importance of reorientational dynamics in hybrid perovskites for their functional performance, many studies have sought to understand the nature of these dynamics, both of simple hybrid halide perovskites and of technologically-relevant complex alloys \cite{christians_tailored_2018}. The remainder of this perspective details several case studies of organic cation dynamics through neutron scattering and complementary techniques.

\section{CASE STUDIES IN HYBRID PEROVSKITE SEMICONDUCTORS}

\begin{figure}[t]
\includegraphics[width=6.3in]{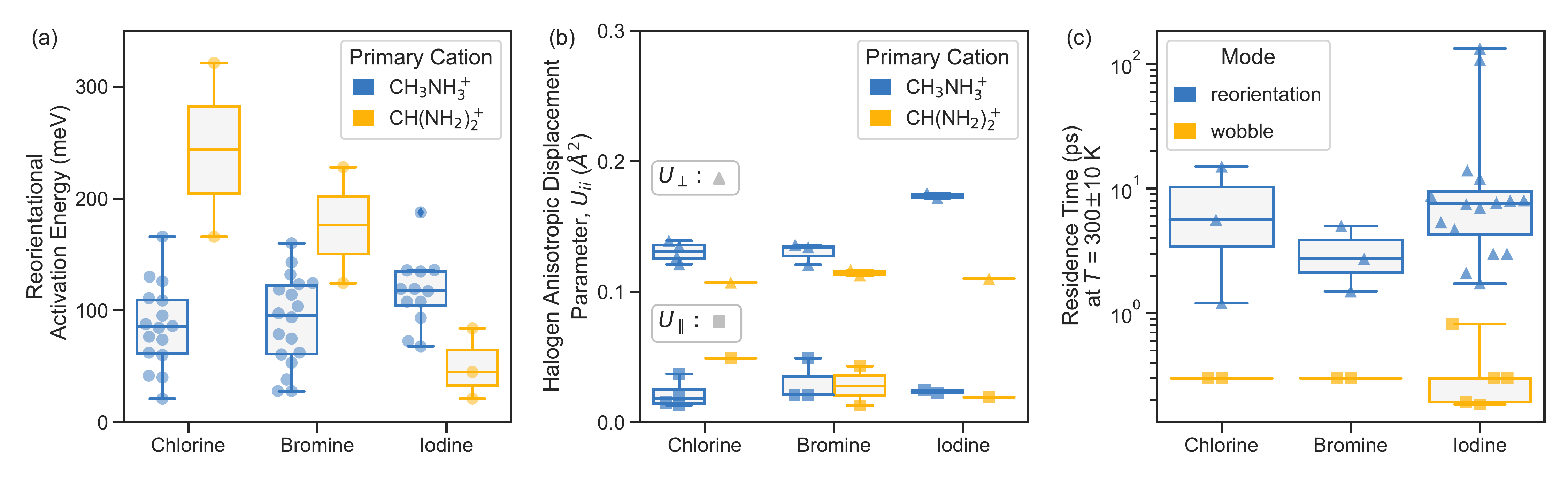}
\caption{
Distribution of activation energies and residence times for organic cation dynamics $A$Pb$X_3$ phases, as differentiated by halogen ($A$ = methylammonium, formamidinium; $X$ = Cl, Br, I).  
(a) Activation energies for molecular reorientational motion further differentiated by organic cation, illustrated by box plots and swarms of individual data points.
(b) Anisotropic displacement parameters (ADPs) of halogen positions in cubic unit cells ($Pm\bar{3}m$)   further differentiated  by organic cation, illustrated by box plots and swarms of individual data points.  The ADPs are separated into those perpendicular to  ($U_\perp$) and parallel to ($U_\parallel$) the Pb-$X$ bonds.  
(c) Residence times at $T = 300\pm10$~K for either reorientational (e.g., whole molecule tumble) or wobble (e.g., $C_3$ rotation about a C-N axis in methylammonium) motions.  This includes methylammonium and formamidinium-based phases in their  tetragonal and cubic phases.  
Box plots illustrate the median (horizontal lines) and mean (box center), while the  markers illustrate data points.  
Data compiled from multiple references are tabulated in the Supporting Information \cite{onoda-yamamuro_dielectric_1992,chen_rotational_2015,bernard_methylammonium_2018,bakulin_real-time_2015,fabini_universal_2017,mozur_dynamical_2019,mozur_orientational_2017,mozur_cesium_2020,selig_organic_2017,swainson_soft_2015,onoda-yamamuro_calorimetric_1990,li_activation_2018,kanno_rotational_2017,xu_molecular_1991, furukawa_cationic_1989,knop_alkylammonium_1990,li_allinorganic_2018}.
}
\label{fig:summary}
\end{figure}

Myriad competing interactions govern the complex dynamics in hybrid perovskites. The activation energies and residence times of the reorientations of  organic cations for several $A$Pb$X_3$ perovskites shown in \textbf{Figure \ref{fig:summary}} reflect several trends in ionic size and hydrogen bonding strength.  In phases where reorientational dynamics are present (e.g., tetragonal and cubic; \textbf{Figure~\ref{fig:summary}(a)}), the activation energy of that motion shows distinct trends for each organic cation. In methylammonium perovskites, the activation energy stays relatively constant as a function of halogen, which correlates more strongly with the constant energy of hydrogen bonding across the halogen series (0.26-0.27 eV/cell) \cite{svane_how_2017} than with the changing size of the halogen (\textbf{Figure~\ref{fig:structure}}).  However, the nature of the halogen does influence the reorientational barrier in formamidinium perovskites, where the increasing size of the halogen leads to an increased cuboctahedral void size to allow for less inhibited molecular reorientations.  This also follows the expected trend in hydrogen bonding strength: 0.16 eV/cell for Cl, 0.1 eV/cell for Br, and 0.09  eV/cell for I \cite{svane_how_2017}.   The smaller size  of methylammonium decreases the importance of the cuboctahedral void size; the stronger dipole moment of methylammonium yields a higher contribution of hydrogen bonding to the total energy \cite{svane_how_2017}.  However, the activation energy for methylammonium reorientations increases for iodine, which correlates with an increase in the crystallographic atomic displacement parameter of iodide  perpendicular to the Pb-$X$ bond that dynamically decreases the cuboctahedral void size (\textbf{Figure~\ref{fig:summary}(b)}).   

Additional complexity is apparent upon examination of the residence times of organic molecules. For both methylammonium and formamidinium perovskites, studies reveal two characteristic residence times for two distinct motions identified as a long time-scale reorientation  and a short time-scale wobble (\textbf{Figure~\ref{fig:summary}(c)}, $\sim17$ ps vs $\sim0.33$ ps, respectively). These data demonstrate the composition dependent interplay of hydrogen bonding and steric interactions on each reorientational mode.

In this section, we present case studies  of organic cation dynamics based on neutron spectroscopy and complementary techniques  (e.g., NMR, dielectric spectroscopy, time-resolved vibrational spectroscopy), as well as their  dependence on chemical substitution, and their relationships with optoelectronic properties.  These case studies demonstrate the  insight that neutron scattering has to offer for hybrid halide perovskites.

\subsection{Unsubstituted Hybrid Halide Perovskite}

\subsubsection{Methylammonium Lead Halide Perovskites}

\begin{figure}[t]
\includegraphics[width=6.3in]{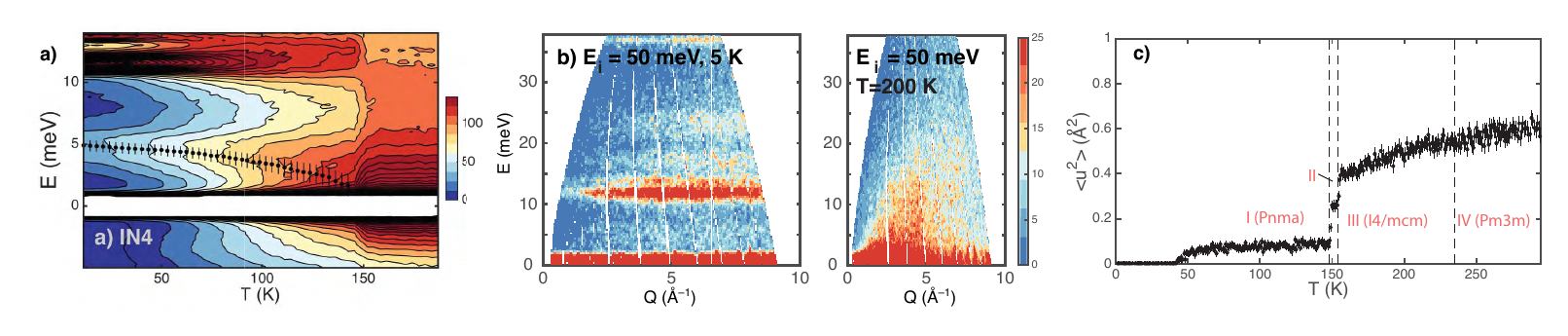}
\caption{ 
(a) Inelastic neutron scattering ($Q$ integrated and powder averaged) collected on the IN4 spectrometer from \ce{CH3NH3PbBr3} showing softening of a $\sim$5 meV phonon as well as the replacement of a $\sim$11 meV vibrational mode with a broad, quasi-elastic relaxation above the orthorhombic to tetragonal phase transition.  
(b) Inelastic neutron scattering as $S(Q,\omega)$ from powder \ce{CH3NH3PbBr3} showing the significant overdampening of modes at $T=$ 200 K.  The sharp, harmonic modes are replaced by quasi-elastic fluctuations extending up to high energies.   
(c) MSD of \ce{CH3NH3PbBr3}, $\langle u^2\rangle$, extracted from an elastic window scan as a function of temperature showing significant changes in hydrogen dynamics at the tetragonal to orthorhombic phase transition, as well as below $T \approx$ 50 K. 
Reprinted figure with permission from Swainson, I. P. Stock, C. Parker, S. F. Van Eijck, L. Russina, M. Taylor, J. W. 2015. From Soft Harmonic Phonons to Fast Relaxational Dynamics in CH$_3$NH$_3$PbBr$_3$. \textit{Phys. Rev. B}, 92 (10), 100303. Copyright 2015 by the American Physical Society.}
\label{fig:Swainson2015}
\end{figure}

Methylammonium lead halide perovskites have been the archetypal hybrid perovskite for solar applications, due to the advantageous band gap of methylammonium lead iodide and the ease of preparation of methylammonium lead iodide compared to the cesium and formamidinium analogues \cite{stranks_metal-halide_2015,dastidar_high_2016,chen_entropy-driven_2016}.  Early  nuclear magnetic resonance (NMR) and calorimetric studies demonstrated that methylammonium is dynamically disordered in the high temperature cubic phase, with increased orientational ordering on cooling that relates to the two known phase transitions \cite{poglitsch_dynamic_1987}. Between 400 K and 100 K, methylammonium lead iodide transitions from the high temperature cubic phase (space group: $Pm\bar{3}m$) to a tetragonal phase (space group: $I4/mcm$) characterized by out-of-phase octahedral rotations when looking perpendicular to the $c$ axis (Glazer notation: $a^0a^0c^-$), and then to an orthorhombic phase (space group $Pnma$) characterized by additional octahedral tilt axes (Glazer notation: $a^-b^+a^-$) \cite{poglitsch_dynamic_1987}. These transitions change the shape and size of the cuboctahedral void where methylammonium sits and thus its dynamical behavior.

The phase behavior of \ce{CH3NH3Pb$X$3} perovskites reflects the presence of  organic-inorganic coupling.  The phonon dispersion measured in the cubic phase of methylammonium lead iodide with high energy resolution inelastic X-ray (HERIX) scattering revealed large amplitude and anharmonic zone-edge rotational instabilities of the octahedra and short-range disorder of the octahedra \cite{beecher_direct_2016}.  This disorder is largest for the hybrids, as opposed to all-inorganic materials \cite{laurita_chemical_2017}.   Subsequent neutron scattering revealed that these instabilities are resolution limited in energy, and thus correspond to static (i.e., with lifetime of $\tau > 36$ ps) tetragonal domains similar to the critical scattering seen from \ce{SrTiO3} \cite{weadock_question_2020}.  First principles calculations reveal that these instabilities are strongly coupled to the librations and reorientations of methylammonium \cite{beecher_direct_2016}. 

In \ce{CH3NH3PbBr3}, there is a clear correlation between the phonons of the Pb-Br framework and methylammonium dynamics.  At the orthorhombic to tetragonal order-disorder transition, sharp modes attributed to methylammonium vibrations decrease in scattered amplitude while heating; at the transition these modes are replaced by broad QENS associated with methylammonium reorientation (\textbf{Figure~\ref{fig:Swainson2015}(a)}).  The decrease in vibrational mode scattering amplitude accompanies the softening of a $\sim$5 meV optical phonon from the Pb-Br framework.  Furthermore, the tetragonal phase phonons become strongly overdamped when the organic cations become significantly dynamic (\textbf{Figure~\ref{fig:Swainson2015}(b)}). In \ce{CH3NH3PbCl3} one observes overdampening of the framework phonon modes across the same temperature scale as observed with bromine, but without the order-disorder phase transition \cite{songvilay_decoupled_2019}.

Given the significance of hydrogen bonding in \ce{CH3NH3Pb$X$_3} \cite{svane_how_2017}, the organic-inorganic interactions are a likely source of anharmonic behavior, as supported by density functional theory calculations of the rotational energy barriers \cite{kanno_rotational_2017}. Organic-inorganic coupling in the form of so called ``nodding donkey'' modes are observed in Raman and infrared spectroscopy experiments and density functional theory-based calculations thereof \cite{brivio_lattice_2015,leguy_dynamic_2016}.  The vacancy ordered double perovskites that lack covalent connectivity between octahedra (e.g., \ce{(CH3NH3)2SnI6}) also show anharmonic tilting behavior that is correlated with the presence of organic cations \cite{maughan_anharmonicity_2018}.  Understanding this anharmonicity, and its origins in the organic cation dynamics, is important for describing locally-broken symmetries and describing the nature of electron-phonon coupling beyond the linear response \cite{whalley_perspective_2017}.

\begin{figure}
\includegraphics[width=6in]{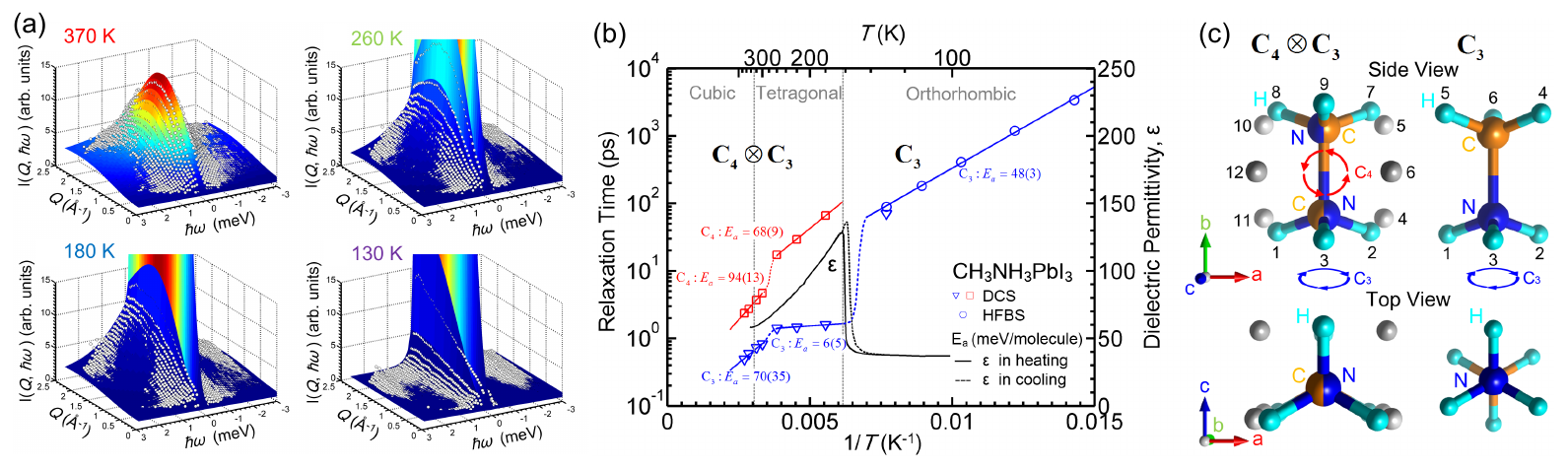}
\caption{(a) QENS from CH$_3$NH$_3$PbI$_3$. Neutron scattering intensity is shown as a function of momentum ($Q$) and energy ($\hbar\omega$) transfers, measured  at 370 K,   260 K,  180 K, and   130 K. Open white circles are the data, and the color coded surface is the surface image of the calculated jump model.
(b) Relaxation time, $\tau$, of rotational modes of CH$_3$NH$_3$PbI$_3$  obtained from fitting QENS data at various temperatures and  plotted against inverse temperature ($1/T$). The colored solid lines are the fits to $\ln(\tau)=E_a /(k_bT)-\ln(A)$, where $E_a$ is an activation energy in units of meV per one molecule, $k_B$ the Boltzmann constant, and $A$ a pre-exponential factor. The colored dotted lines are guides to the eye. The black solid and dotted lines are temperature dependent dielectric permittivity measured at 1 kHz by Onoda-Yamamuro, et al. \cite{onoda-yamamuro_dielectric_1992}, which show a sharp increase at 160 K. 
(c) $C_3 \otimes C_4$: illustration of 12 crystallographically-equivalent sites for H atoms of CH$_3$NH$_3^+$ in the tetragonal environment. $C_3$: 3 equivalent H atomic sites in the orthorhombic environment. Reprinted figure with permission from T. Chen, B. Foley, B. Ipek, M. Tyagi, J. R. D. Copely, C. M. Brown, J. J. Choi, S. H. Lee. 2015. \textit{Physical Chemistry Chemical Physics} 17 (46) 31278-31286. Copyright Royal Society of Chemistry 2015.}
\label{fig:Chen2015}
\end{figure}

Methylammonium rapidly reorients in the high-temperature cubic phase and exhibits a decrease in amplitude, timescale, and dynamic degrees of freedom on cooling as the cuboctahedral void constricts, as learned from  extensive neutron scattering  and complementary NMR spectroscopy studies.  In methylammonium lead bromide, the temperature-dependent mean squared displacements (MSDs) calculated from incoherent elastic neutron scattering   (\textbf{Figure \ref{fig:Swainson2015}(c)}, \cite{swainson_soft_2015}) capture the overall decrease in motion. The lack of anomalies in $\langle u^2\rangle$ at the cubic to tetragonal phase transition suggest that the entropy loss associated with this phase transition  (8.2 J mol$^{-1}$ K$^{-1}$, $\approx R \ln(2.7)$ \cite{onoda-yamamuro_calorimetric_1990}) correlates to a change in the type of motion rather than a reduction in the spatial extent of methylammonium dynamics, $\langle u^2\rangle$. In contrast, the clear feature in the MSD at the tetragonal to orthorhombic phase transition suggests a discontinuous change in how methylammonium reorients in the cuboctahedral void.  These observations are also consistent with the conservation of \ce{^1H} and \ce{^{14}N} NMR lineshapes across the cubic to tetragonal phase transition and appreciable narrowing after the tetragonal to orthorhombic phase transition \cite{xu_molecular_1991,bernard_methylammonium_2018}.

The reorientational dynamics of methylammonium are well described by a jump model of transitions between idealized orientations, as originally proposed from changes in degrees of freedom at various phase transitions determined from calorimetry \cite{onoda-yamamuro_calorimetric_1990}.  Rigorous group theoretical analysis by Chen, et al. with a jump model provided a complete description of the QENS data (\textbf{Figure~\ref{fig:Chen2015}}, \cite{chen_rotational_2015}).  This jump model involves the direct product of a $C_3$  rotation (120$^\circ$) derived from the internal symmetry axis of methylammonium and a $C_4$ rotation (90$^\circ$)  perpendicular to the $C_3$ rotation as imposed by the symmetry of the lattice site point group symmetry. Both the $C_3$ and $C_4$ modes are active in the cubic and tetragonal phases, with relaxation times of $\sim1$ ps and $\sim5$ ps, respectively, as determined by two different studies \cite{chen_rotational_2015, li_activation_2018}.  This analysis of the QENS is consistent with direct measurements of the molecular dynamics; two-dimensional transient IR spectroscopy also reveals the presence of a fast (300 fs) wobbling-in-a-cone motion similar to the $C_3$ rotation and a slow (3 ps) molecular reorientation similar to the $C_4$ rotation \cite{bakulin_real-time_2015}.

The molecular reorientations correlate most strongly to the dielectric response.  As shown in \textbf{Figure~\ref{fig:Chen2015}(b)}, the $C_4$ motion increases significantly in relaxation time on cooling within the tetragonal phase, but this reorientation is frozen out upon cooling into the orthorhombic phase.  Therefore, loss of the $C_4$ reorientation leads to a significant decrease in the dielectric function \cite{onoda-yamamuro_dielectric_1992, govinda_behavior_2017, govinda_critical_2018, anusca_dielectric_2017, poglitsch_dynamic_1987}, schematically depicted in \textbf{Figure~\ref{fig:neutron}}.  

Through this organic-inorganic coupling, the dynamic degrees of freedom of methylammonium correlate to changes in the optoelectronic properties. For example, the freezing out of the $C_4$ whole molecular reorientation at the tetragonal-orthorhombic phase transition correlates to faster recombination of excited state charge carriers as well as an increase in exciton binding energies \cite{hutter_directindirect_2017, miyata_direct_2015}.  These correlations suggest that the methylammonium $C_4$ reorientations participate in the electron-phonon models described in \textbf{Figure \ref{fig:structure}}.

\subsubsection{Formamidinium Lead  Perovskite Halides}

\begin{figure}[t]
\includegraphics[width=6in]{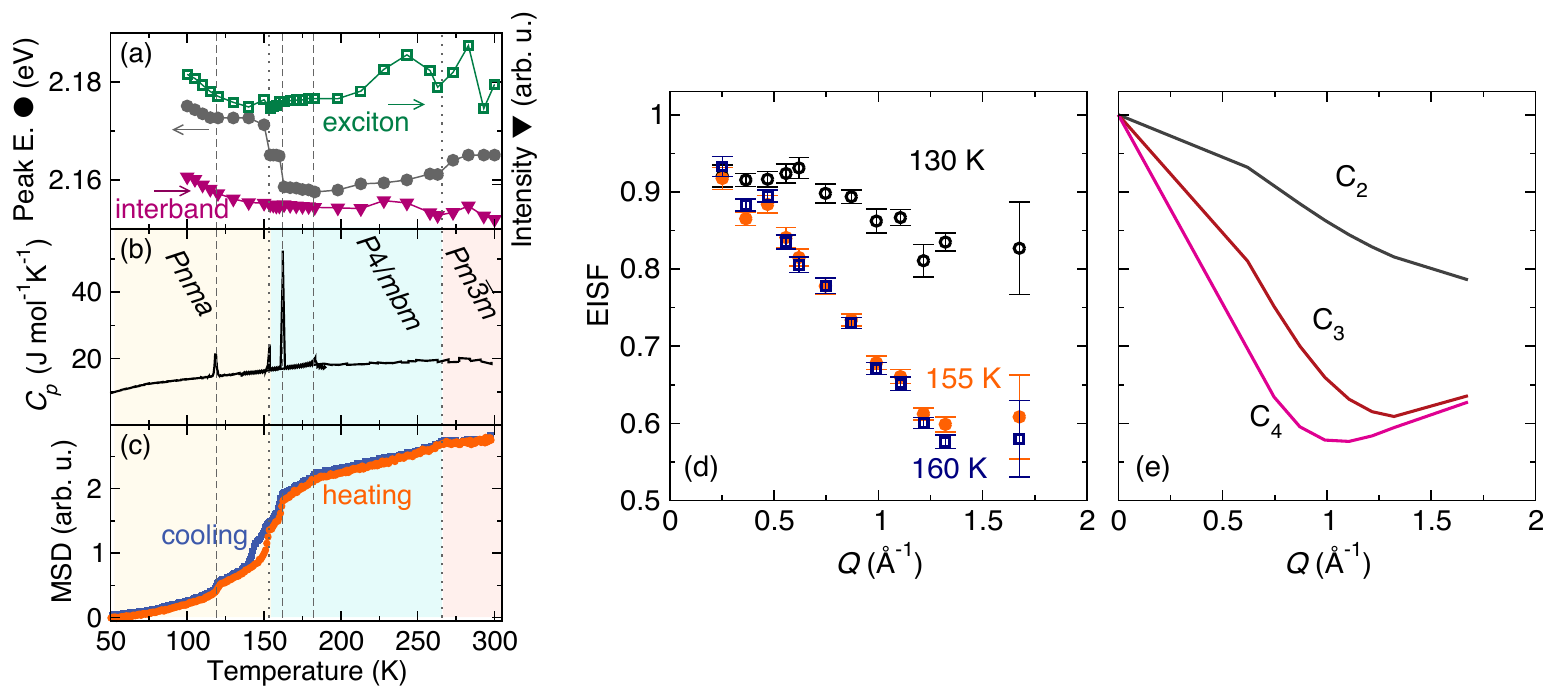}
\caption{
(a) Parameters from modelling the temperature-dependent photoconductivity of CH(NH$_2$)$_2$PbBr$_3$. Excitonic peak center (left axis, gray circles), excitonic peak intensity (right axis, green squares), and interband intensity (right axis, pink triangles). (b) Heat capacity data for CH(NH$_2$)$_2$PbBr$_3$. The crystallographically-observed phase transitions occur at 266 K and 153 K. (c) MSDs determined from fixed window elastic neutron scattering  of CH(NH$_2$)$_2$PbBr$_3$ measured on heating (orange circles) and cooling (blue squares). The colored, rectangular overlays illustrate the cyrstallographic symmetry at each temperature, with dotted gray lines indicating the crystallographic phase transition temperatures and dashed gray lines showing other phase transition temperatures determined from heat capacity and differential scanning calorimetry data. (d) EISFs extracted from QENS spectra at T = 160 K, 155 K, and 130 K. For comparison, models of $C_2$, $C_3$, and $C_4$ molecular rotations are shown in (e). 
Reprinted with permission from Mozur, E. M., Trowbridge, J. C., Maughan, A. E., Gorman, M. J., Brown, C. M., Prisk, T. R., Neilson, J. R. 2019. Dynamical Phase Transitions and Cation Orientation-Dependent Photoconductivity in \ce{CH(NH2)2PbBr3}. \textit{ACS Mater. Lett}. 2019, 1 (2), 260–264. Copyright 2019 American Chemical Society.
}
\label{fig:Mozur2019}
\end{figure}

Formamidinium lead halide perovskites typically have unusual phase behavior compared to methylammonium and cesium analogues, including reentrant phase transitions \cite{fabini_reentrant_2016}, path dependent phase transitions \cite{chen_entropy-driven_2016}, and crystallographically unresolvable phase transitions \cite{mozur_dynamical_2019}.
Similar to methylammonium perovskites, formamidinium lead halide perovskites exhibit dynamic degrees of freedom within the organic sublattice that freeze out with temperature, which is reflected in the temperature dependence of the dielectric permittivity \cite{fabini_reentrant_2016, fabini_universal_2017, govinda_critical_2018, mozur_dynamical_2019}. Anomalies  in the temperature-dependent permittivity also occur at phase transition temperatures. Despite these similarities with methylammonium perovskites, the distinct molecular shape and delocalized cationic $\pi$-system yield several distinctions in formamidinium perovskites. 

Formamidinium lead bromide is notable for having five temperature-dependent anomalies in the MSDs that correlate with anomalies in the heat capacity and  in the photoconductivity (\textbf{Figure \ref{fig:Mozur2019}}  \cite{mozur_dynamical_2019}), as well as the thermal expansion from dilatometry \cite{keshavarz_tracking_2019}.  Two of these phase transitions correspond to changes in crystallographic symmetry with detectable changes in the octahedral tilting and rotation patterns \cite{schueller_crystal_2018, mozur_dynamical_2019}, while three phase transitions release entropy they do not correlate with a readily detectable change in crystallographic symmetry \cite{mozur_dynamical_2019, mozur_cesium_2020, keshavarz_tracking_2019}. Instead, they are characterized by changes in the dynamic degrees of freedom of formamidinium. At the time of this writing, these crystallographically-unresolvable phase transitions are unique among hybrid perovskites, although formamidinium lead iodide shows unusual, sample history dependent phase transitions that are either reentrant and/or entropy stabilized \cite{chen_entropy-driven_2016, fabini_reentrant_2016}.

Based on the molecular geometry, the natural rotations for formamidinium are  180$^\circ$ rotations with the axis of rotation parallel to the carbon-hydrogen bond or about the axis connecting the two nitrogen atoms, as well as a 120$^\circ$ rotation orthogonal to the molecular plane (\textbf{Figure~\ref{fig:structure}(a)}).  The point group symmetry of the ideal lattice site occupied by formamidinium suggests that the molecule could undergo reorientational $C_4$ rotations.  Molecular dynamics simulations and density functional theory calculations  predict that $C_2$ rotations, where the axis of rotation connects the two nitrogen atoms rather than a higher symmetry location, are the lowest energy molecular dynamics \cite{carignano_close_2016, fabini_universal_2017}; these simulations also suggest the presence of a slow reorientation and a fast reorientation, in analogy to those observed in methylammonium perovskites \cite{carignano_close_2016}.  However, direct modeling of the experimental EISF is complicated by the complex orientational potential energy landscape of formamidinium within the nearly cuboctahedral void.  The EISF resembles a motion between idealized $C_3$ (120$^\circ$) and $C_4$ (90$^\circ$) rotations at elevated temperatures but 180$^\circ$ rotation at lower temperature  (\textbf{Figure~\ref{fig:Mozur2019}(e)}), consistent with changes in the relaxation times marked by the reduction in QENS peak width across phase transitions \cite{mozur_dynamical_2019}.

The complex behavior of formamidinium dynamics appears to be related to its distinct distribution of electrostatic charges. Formamidinium has a weaker dipole moment and stronger quadrupole moment than methylammonium \cite{chen_origin_2017}.   A pair of molecular quadrupoles embedded in a dielectric crystal prefer to orient perpendicular to each other in a ``T'' shape \cite{grannan_low-temperature_1990},  which can couple strongly to  compressive and expansive strains from the surrounding anionic lattice \cite{mozur_cesium_2020}.  This pairwise ``T'' shape  cannot tile any Bravais lattice, and therefore the molecular orientations of formamidinium are frustrated \cite{mozur_dynamical_2019}.  Frustrated orientations are consistent with NMR results that show that formamidinium reorients more quickly than methylammonium and with dielectric susceptibility measurements that indicate a series of complex changes in formamidinium dynamics at low temperatures \cite{kubicki_cation_2017, schueller_crystal_2018}. This geometric frustration participates in the orientational phase transitions of formamidinium lead bromide observed with QENS and could explain the path dependent phase transitions of formamidinium lead iodide. $^{79}$Br nuclear quadrupolar resonance (NQR) data suggest that the reorientations of formamidinium couple to changes in the local environment around the inorganic octahedra, which aligns with first principle calculations that show that the rotational energy barrier of formamidinium can be tuned with substitution on the halide and metal sites \cite{mozur_cesium_2020, kanno_theoretical_2017}. Understanding the unusual phase behavior of formamidinium perovskites is essential to understanding their properties, as each phase has a different optoelectronic behavior \cite{fabini_reentrant_2016, mozur_dynamical_2019}.

\begin{marginnote}[]
\entry{Nuclear Quadrupolar Resonance (NQR)}{a nuclear resonance technique performed in the absence of a magnetic field on solids containing quadrupolar nuclei, wherein the nuclear spin levels are split by the local chemical environment and interrogated with radio waves.}
\end{marginnote}

\subsubsection{Lead-free Hybrid Perovskite Halides} 

While more difficult to experimentally study (e.g., due to air sensitivity),  tin and germanium perovskites offer new opportunities for functional properties, as they show an increased contribution of the $ns^2$ lone pair electrons and lack the toxicity of lead \cite{yang_tin_2020, ganose_beyond_2017, fabini_underappreciated_2020}. The increased stereochemical activity in tin and germaninum perovskites relative to lead perovskites \cite{laurita_chemical_2017,stoumpos_hybrid_2015} is expected to couple to the dynamic degrees of freedom of methylammonium \cite{remsing_new_2020}, although studies so far have focused predominantly on ionic conductivity \cite{yamada_phase_1998,yamada_static_2002}.  However, density functional theory calculations also indicate that tin and germanium perovskites have higher energy barriers for methylammonium reorientations, attributed to a decrease in lattice polarizability and compression of the $A$-site void \cite{kanno_rotational_2017, herz_how_2018, guedes-sobrinho_thermodynamic_2019}. Furthermore, the overall phase behavior appears to be more sensitive to the molecular dynamics, as isotope enrichment can completely change (e.g., \ce{CH3NH3SnBr3} \cite{swainson_orientational_2010} vs \ce{CD3ND3SnBr3} \cite{onoda-yamamuro_neutron-diffraction_1995}) or suppress phase transitions (e.g., \ce{CH3NH3GeCl3}  vs \ce{CD3ND3GeCl3} \cite{yamada_static_2002}). Experimental validation of these computational predictions and investigation into lone-pair coupling to methylammonium dynamic degrees of freedom provide exciting avenues for continued research.

\subsection{Effects of Chemical Substitution}

Most of the technologically-relevant hybrid halide perovskites for photovoltaic applications are heavily chemically substituted (i.e., alloyed), as exemplified by a high-performing composition: \ce{(CH(NH2)2)_{0.79}(CH3NH3)_{0.16}Cs_{0.05}Pb(I_{0.84}Br_{0.17})3} \cite{christians_tailored_2018}.  
Chemical substitution is easily accomplished synthetically in the solution-phase or solid-state routes to these materials \cite{rosales_synthesis_2018, christians_tailored_2018}, which allows for tuning of optoelectronic properties such as the band gap and polaron mobility \cite{liu_bimodal_2019,zhou_cation_2019}. Furthermore, the unsubstituted hybrid perovskites methylammonium lead iodide and formamidinium lead iodide are enthalpically unstable \cite{nagabhushana_direct_2016} and decompose rapidly under ambient conditions \cite{askar_multinuclear_2017,senocrate_thermochemical_2019,aziz_understanding_2020,cordero_stability_2019};  certain substitutions stabilize the perovskite phases \cite{schelhas_insights_2019}  or slow this decomposition process \cite{christians_tailored_2018, poorkazem_compositional_2018, chen_toward_2019}.  While chemical substitution can improve phase stability, it modifies the internal energy landscape of hybrid perovskites that influences the atomistic dynamics and thus electron-phonon coupling \cite{leppert_electric_2016, zhou_cation_2019}.  

This section focuses on the role of cesium substitution in the two perovskite materials described above, methylammonium lead bromide and formamidinium lead bromides. In contrast to the hybrid materials, cesium lead bromide exhibits orthorhombic symmetry with octahedral tilting at and below room temperature, with the cubic and tetragonal phases accessible at temperatures above 80 $^\circ$C and 140 $^\circ$C, respectively \cite{Rodova_phase_2003}. The generalized phase behavior of the substituted alloys can be understood within the context of a pseudo-binary phase diagram, where compositions closest to the end members behave as solid-solutions separated by a two-phase region for intermediate compositions.  However, the detailed nature of the phase transitions requires closer examination of the dynamic nature of the organic cations in these plastic crystals.  

\subsubsection{Cesium Substitution in Methylammonium Lead Bromide}

\begin{figure}[t]
\includegraphics[width=5in]{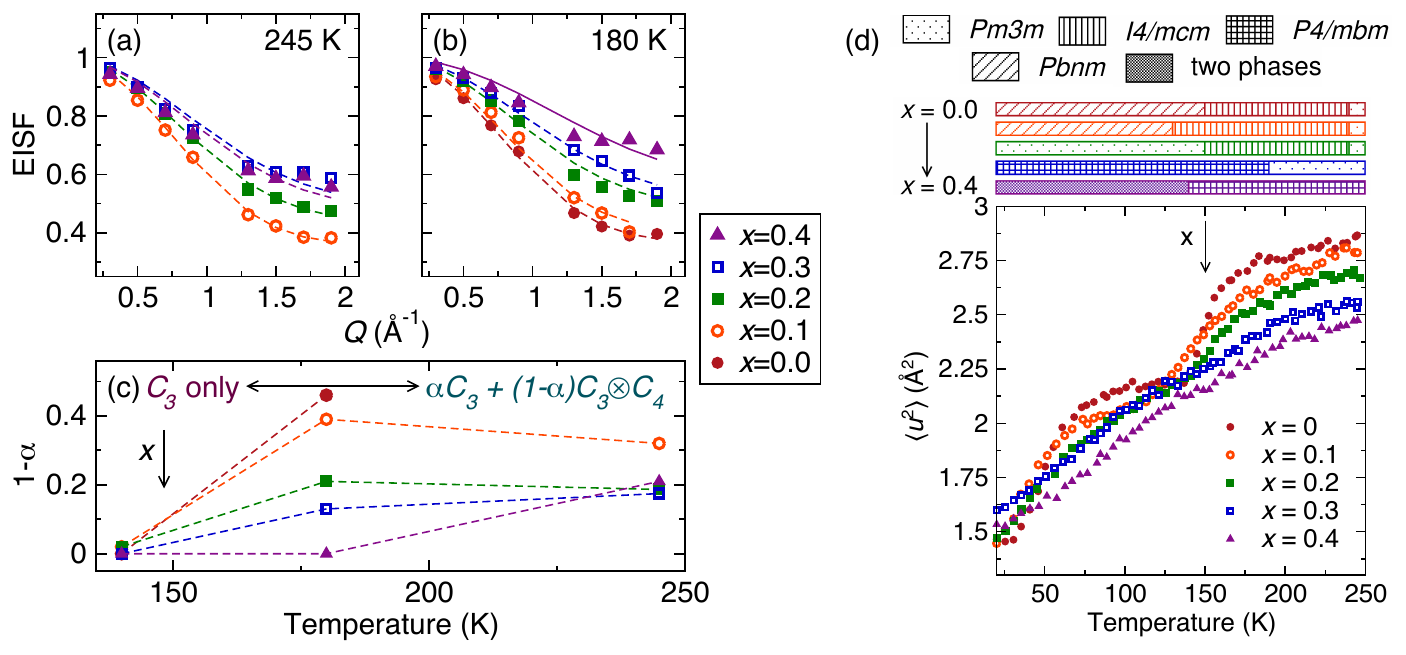}
\caption{Selected EISF values for (CH$_3$NH$_3$)$_{1-x}$Cs$_x$PbBr$_3$ extracted from QENS spectra collected at (a) 245 K and (b) 180 K. Dashed lines are fits to the coupled $C_3$ and $C_4$ jump model, while solid lines are fits to the $C_3$ jump model. (c) The fraction of CH$_3$NH$_3^+$ participating in a $C_3\otimes C_4$ rotation ($1 - \alpha$) from the EISF modeling. Dashed lines are guides to the eye. (d) MSDs as a function of temperature extracted from QENS spectra of (CH$_3$NH$_3$)$_{1-x}$Cs$_x$PbBr$_3$ illustrating a smearing of transitions with increasing $x$. The bars at the top illustrate the observed lattice symmetry of (CH$_3$NH$_3$)$_{1-x}$Cs$_x$PbBr$_3$ for $x$ = 0.0, 0.1, 0.2, 0.3, and 0.4.  Reprinted with permission from Mozur, E. M. Maughan, A. E. Cheng, Y. Huq, A. Jalarvo, N. Daemen, L. L.; Neilson, J. R. 2017. Orientational Glass Formation in Substituted Hybrid Perovskites. \textit{Chem. Mater}. 29 (23), 10168–10177. Copyright 2017 American Chemical Society.}
\label{fig:Mozur2017}
\end{figure}

In methylammonium lead bromide, cesium substitution for methylammonium leads to inhibition of the methylammonium dynamics. As discussed above, methylammonium undergoes 90$^\circ$ rotations perpendicular   and 120$^\circ$ rotations parallel to the carbon-nitrogen bond axis \cite{chen_rotational_2015}. Substitution of cesium for methylammonium in the solid-solution, (CH$_3$NH$_3$)$_{1-x}$Cs$_x$PbBr$_3$, does not affect the identity of these rotations; however, modeling of the QENS spectra demonstrates that a fraction of molecules become pinned as the concentration of cesium increases (\textbf{Figure~\ref{fig:Mozur2017}(a-c)}) \cite{mozur_orientational_2017}.  As discussed previously, the molecular reorientations are directly responsible for a significant fraction of the increased dielectric polarizability (\textbf{Figure~\ref{fig:Chen2015}(b)}).

The pinning of molecular reorientations with cesium substitution in (CH$_3$NH$_3$)$_{1-x}$Cs$_x$PbBr$_3$ correlates with a significant modification of the phase behavior.  In contrast to \ce{CH3NH3PbBr3} (\textbf{Figure~\ref{fig:Swainson2015}}), the temperature dependent MSDs vary smoothly with temperature and do not show abrupt changes near crystallographic phase transition temperatures in the substituted materials (\textbf{Figure~\ref{fig:Mozur2017}}) \cite{mozur_orientational_2017}.  With increasing substitution, crystallography indicates that the phase transitions occur over a larger temperature range and are dependent on temperature ramp rate \cite{mozur_orientational_2017}. The sluggish phase transitions and inhibited dynamics indicate that methylammonium lead bromide becomes an orientational glass upon cesium substitution characterized by disordered and frozen methylammonium orientations, similar to bromide substitution in potassium cyanide \cite{loidl_orientational_1989}.

In analogy to spin glasses, the formation of an orientational glass relates to frustration of order parameters related to the orientation of the molecule \cite{loidl_orientational_1989}. One proposed source of the increased frustration on substitution is that cesium and methylammonium prefer different coordination from the octahedral framework.  Most halide perovskites crystallize in the tetragonal $P4/mbm$ phase, characterized by in-phase octahedral rotations, as this configuration maximizes the $A$-site coordination \cite{young_octahedral_2016}. Hydrogen-bonding interactions between methylammonium and the octahedral framework create a preference of the tetragonal $I4/mcm$ phase for methylammonium perovskites, characterized by out-of-phase octahedral rotations \cite{lee_role_2015,franz_interaction_2016, yin_hydrogen-bonding_2017}. These interactions, combined with the contraction of the $A$-site with cesium substitution, compete and lead to the inhibition of methylammonium reorientations and the formation of an orientational glass.  It appears that substitution increases the influence of hydrogen bonding (cf.,  \textbf{Figure~\ref{fig:summary}}).  
Recent first principles calculations suggest that reductions in symmetry, such as those from frozen or inhibited methylammonium reorientations, change the binding energy of polarons \cite{zhou_cation_2019} or tune the magnitude of Rashba splitting \cite{leppert_electric_2016}, which suggests that substitution impacts electronic properties through the organic cation dynamics.  

\subsubsection{Cesium Substitution in Formamidinium Lead Bromide}

\begin{figure}[t]
\includegraphics[width=6in]{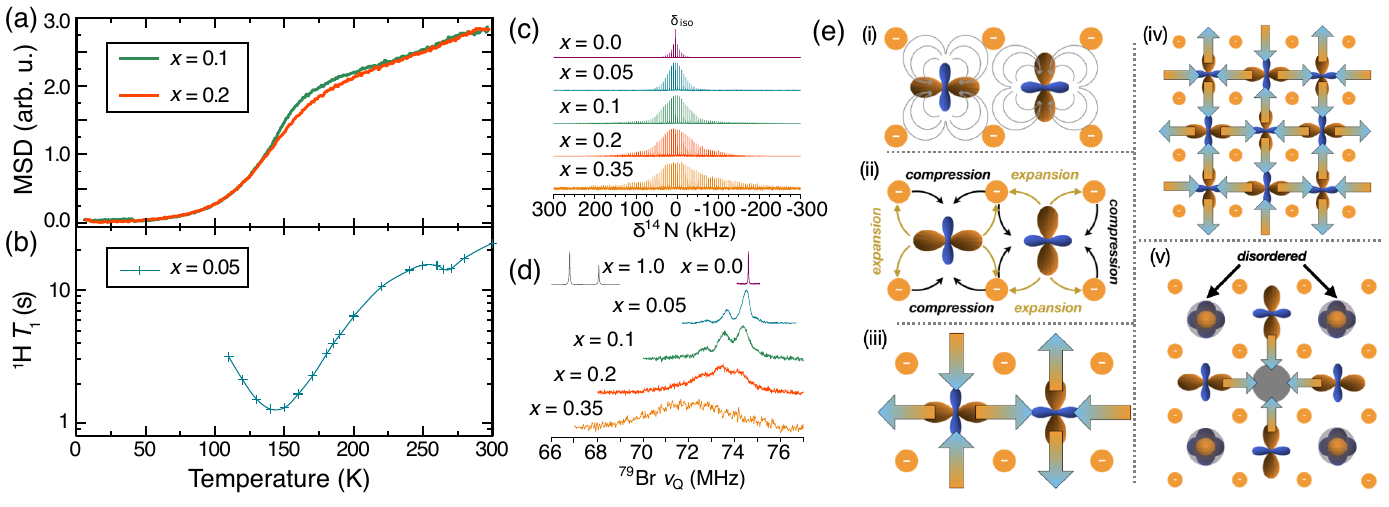}
\caption{
(a) MSDs extracted from fixed window elastic neutron scattering spectra for (CH(NH$_2$)$_2$)$_{1-x}$Cs$_x$PbBr$_3$, where $x$ = 0.1 and 0.2 illustrating a lack of discontinuities in substituted samples, in contrast to Figure~\ref{fig:Mozur2019}(c). 
(b) Spin-lattice relaxation time ($T_1$) determined from $^1$H NMR spectroscopy of  (CH(NH$_2$)$_2$)$_{0.95}$Cs$_{0.05}$PbBr$_3$ illustrating one phase transition at $T\sim$ 260 K.  
(c) Room-temperature \ce{^{14}N} NMR spectra of (CH(NH$_2$)$_2$)$_{1-x}$Cs$_x$PbBr$_3$ where $x$= 0 to 0.35 illustrating an increased anisotropy of cation rotations with $x$. 
(d) Room-temperature \ce{^{79}Br} NQR spectra collected on (CH(NH$_2$)$_2$)$_{1-x}$Cs$_x$PbBr$_3$ showing how increasing the level of substitution leads to an effectively continuous distribution of possible local environments.
(e) Cartoon illustrating electrostatic potential field lines emanating from the quadrupolar moment of formamidinium (orange density denotes negative; blue is positive) yield attractive or repulsive interactions to the anionic framework, as depicted in (i) and (ii). (iii) The resulting expansive and compressive strain fields map onto an elastic dipole tensor (large arrows). (iv) In a plane, the elastic dipoles can order; however, a favorable 3D configuration cannot be tiled. (v) The smaller \ce{Cs+} cation introduces compressive local strain, thus causing a preferred orientation of quadrupolar cations around the \ce{Cs+} and frustrated orientations of the next-nearest neighboring quadrupoles.
Adapted with permission from Mozur, E. M. Hope, M. Trowbridge, J. C. Halat, D. M. Daemen, L. L. Grey, C. P. Neilson, J. R. 2020. Cesium Substitution Disrupts Concerted Cation Dynamics in Formamidinium Hybrid Perovskites.\textit{Chem. Mater.} 32, (14) 6266–6277 Copyright 2020 American Chemical Society.}
\label{fig:Mozur2020}
\end{figure}

The orientational phase transitions of formamidinium lead bromide are sensitive to chemical substitution. The five phase transitions of formamidinium lead bromide have dynamic signatures that manifest in \ce{^1H} NMR, \ce{^{14}N} NMR, and MSDs between room temperature and 100 K (\textbf{Figure~\ref{fig:Mozur2019}}).  However, formamidinium lead bromide substituted with as little as 5-10\% cesium exhibits only one anomaly in \ce{^1H} NMR, \ce{^{14}N} NMR, or in MSDs between room temperature and 100 K, which  corresponds to the high-temperature cubic to tetragonal phase transition  (\textbf{Figure \ref{fig:Mozur2020}(a)}) \cite{mozur_cesium_2020}.

Cesium substitution effectively suppresses the four low temperature phase transitions by interrupting the concerted changes in formamidinium dynamics.  At room temperature, \ce{^{14}N} NMR spectra of (CH(NH$_2$)$_2$)$_{1-x}$Cs$_x$PbBr$_3$ show retained reorientational dynamics, but with an increased anisotropy of the \ce{^{14}N} environment with $x$; this accompanies a clear distortion of the local \ce{^{79}Br} environments, as determined from NQR spectroscopy (\textbf{Figure~\ref{fig:Mozur2020}(b)}).  The interplay between substitution and formamidinium orientation reflects a distinct behavior from methylammonium (\textbf{Figure~\ref{fig:summary}}), which can be linked to the large electrostatic quadrupolar moment of formamidinium, as depicted in  \textbf{Figure~\ref{fig:Mozur2020}(c)}.  The relative orientations of formamidinium within the inorganic framework are mediated by the electrostatic quadrupolar field, which is very sensitive to the surrounding electric field gradient.  When this interaction is mapped onto an elastic dipole tensor, it is possible to see how a stable long-range ordered configuration becomes challenging to produce (i.e., it is frustrated).  Introduction of compressive microstrain from the smaller cesium cation imparts a significant change to the potential energy landscape with consequences to the long range order and overarching phase transitions \cite{mozur_cesium_2020}.  While the long-range order is strongly influenced by the substitution, the local dynamics of formamidinium appear to be mostly retained \cite{mozur_cesium_2020}, which suggest that formamidinium-based perovskites serve as a better host for substituted alloys if one wishes to maintain a larger  dielectric response.

\subsubsection{Complex, Multiple Substitutions in Hybrid  Perovskite Halides}

The ability to use a cocktail of cations and anions to create highly stable, functional photovoltaic devices reflects a unique defect tolerance of hybrid perovskites.\cite{christians_tailored_2018}.  While pseudo-binary solid-solutions formed by substitution on either $A$-site  (e.g., (CH(NH$_2$)$_2$)$_{1-x}$Cs$_x$PbI$_3$ \cite{li_microscopic_2020}) or the halogen site  (e.g., \ce{CH3NH3Pb(I_{$1-x$}Br_$x$)3} \cite{slotcavage_light-induced_2016} )
are prone to phase separation and decomposition, complex, multiple substitutions on both $A$ and halogen sites prevents this phase separation \cite{correa-baena_homogenized_2019}.  The chemical pressure from  cesium substitution for methylammonium in \ce{(CH3NH3)_{0.5}Cs_{0.5}Pb(Br_{0.5}I_{0.5})3} appears to have the same influence as exerting 0.2 GPa of external pressure on \ce{CH3NH3Pb(Br_{0.5}I_{0.5})3} and thermodynamically stabilizes the solid solution, highlighting the importance of an ``often-overlooked'' $P\Delta V$ term in the thermodynamic potential \cite{hutter_thermodynamic_2020}.   

Therefore, one expects complex substitutions to also influence the organic molecular dynamics.  The inhibition with cesium substitution in methylammonium perovskites resembles methylammonium-substituted formamidinium lead iodide, in which  lineshape analysis of \ce{^{14}N} NMR spectra reveals that methylammonium and formamidinium reorientations become increasingly anisotropic with increased substitution \cite{kubicki_cation_2017}. 
Substitution at the halide site, as used to increase stability and tune the electronic bandgap \cite{lee_efficient_2012,christians_tailored_2018}, also leads to inhibited cation dynamics. Pump-probe two-dimensional infrared spectroscopy  studies  of \ce{CH3NH3Pb(Cl_$x$Br_{$1-x$})3} and \ce{CH3NH3Pb(Br_$x$I_{$1-x$})3} show that both series exhibit slower methylammonium dynamics than either end member \cite{selig_organic_2017}.  This supports the observation of decreased orientational disorder in methylammonium lead bromide after chlorine substitution, as determined by Fourier difference maps from Rietveld refinement of neutron powder diffraction \cite{lopez_dynamic_2019}.  Many of the top-performing photovoltaic devices substitute a formamidinium-based host \cite{kim_high-efficiency_2020}.  Given that the local formamidinium dynamics are not significantly inhibited by substitution \cite{mozur_cesium_2020}, any of the advantages provided by having a dynamic organic cation are still active in formamidinium-based materials.

\section{CONCLUSIONS AND OUTLOOK}
Collective and local dynamics in hybrid perovskites underlie the optoelectronic properties that make them of interest for photovoltaics, radiation detection, and light emitting diodes. Therefore, characterization of these dynamics is essential for building useful theories to describe their behavior, especially for local dynamics that do not yet have rigorous models. Neutron scattering plays a large role in uncovering these dynamics -- the dynamic structure factor, $S(Q,\omega)$, informs researchers of the correlations of these molecular dynamics in both space and in time.  Complementary methods, such as NMR, dielectric spectroscopies, X-ray scattering, molecular dynamics simulations, and DFT calculations, provide a complete perspective across many environmental conditions. Neutron scattering has been used to track the changes in dynamics as a function of temperature, to describe the symmetry of molecular reorientations, and to  characterize their activation energies. These investigations demonstrate that these local dynamics are affected by cation identity, electrostatic and hydrogen-bonding interactions with the extended octahedral framework, and substitution.  This creates a deeper understanding of the phase behavior of hybrid halide perovskites.  Investigations into 2D perovskite materials have only just begun, and many questions remain about the dynamic degrees of freedom of bulky organic cations and their relationship to optoelectronic properties. Similar questions remain for other hybrid perovskite derivatives, including zero dimensional vacancy ordered perovskites. These insights are the basis of rigorous structure-property relationships that will enable the rational innovation with next-generation semiconductors.

\begin{summary}[SUMMARY POINTS]
\begin{enumerate}
\item Hybrid perovskite semiconductors are plastic crystals with dynamically reorienting organic cations.  
\item Organic cation dynamics have a direct influence on the dielectric response of hybrid perovskite semiconductors.  
\item Molecular reorientational dynamics exert indirect influences on the optical and electronic properties of hybrid perovskites by way of locally broken symmetry, electron-phonon coupling, and defect creation and motion. 
\item Chemical substitution of hybrid perovskite semiconductors influences the thermodynamic stability and organic cation dynamics.  
\end{enumerate}
\end{summary}

\section*{DISCLOSURE STATEMENT}
The authors are not aware of any affiliations, memberships, funding, or financial holdings that
might be perceived as affecting the objectivity of this review. 

\section*{ACKNOWLEDGMENTS}
This work was supported by the U.S. Department of Energy, Office of Science, Basic Energy Sciences, under Award SC0016083. J.R.N. and E.M.M also acknowledge support from Research Corporation for Science Advancement through a Cottrell Scholar Award and the A.P. Sloan Foundation for assistance provided from a Sloan Research Fellowship. The authors also thank K.~A.~Ross and A.~E.~Maughan for insightful discussions.

\bibliographystyle{ar-style3}

\bibliography{references.bib}

\end{document}